\documentclass[aps,twocolumn,showpacs,preprintnumbers,noshowpacs,superscriptaddress,6pt]{revtex4-1}
\usepackage[labelfont=bf]{caption}
\usepackage{euscript}
\usepackage{epsf}
\usepackage{epsfig}
\usepackage{subcaption}
\usepackage[english]{babel} 
\usepackage[utf8]{inputenc}
\usepackage{graphicx,epstopdf}
\usepackage{overpic}
\usepackage{multirow}
\usepackage{adjustbox}
\usepackage{natbib}

\usepackage{parskip}
\usepackage[T1]{fontenc}
\usepackage{dcolumn}

\usepackage{bm}
\usepackage{amsfonts}
\usepackage{amsmath}
\usepackage{amssymb}
\usepackage{siunitx}
\usepackage{booktabs}
\usepackage{mathtools}
\usepackage{flafter}
\usepackage{threeparttable}
\usepackage{subfiles}
\usepackage{import}
\usepackage{float}
\usepackage{makecell}
\usepackage{diagbox}
\usepackage{braket}
\usepackage{indentfirst}
\usepackage{mdframed}
\usepackage[strict]{chngpage}

\newmdenv{allfour}

\newmdenv[leftline=false,rightline=false]{topbot}

\newmdenv[topline=false,rightline=false]{leftbot}

\usepackage{ulem}

\usepackage{color, colortbl}

\definecolor{Gray}{gray}{0.9}
\definecolor{LightCyan}{rgb}{0.88,1,1}

\definecolor{BLUE}{rgb}{0.0,0.0,1.0}

\renewcommand\[{\begin{equation}}
\renewcommand\]{\end{equation}}
\graphicspath{Figs/}

\everymath{\displaystyle}
\renewcommand{\vec}[1]{{\mbox{\boldmath$#1$}}}

\begin{document}

\title{
A relativistic coupled-cluster treatment of magnetic hyperfine structure of the $X^2\Pi$ and $A^2\Sigma^+$ states of OH isotopologues
}

\author{D.~P.~Usov}
\affiliation{Department of Physics, St. Petersburg State University, 7/9 Universitetskaya nab., 199034 St. Petersburg, Russia}

\author{Y.~S.~Kozhedub}
\email[]{y.kozhedub@sbpu.ru}
\affiliation{Department of Physics, St. Petersburg State University, 7/9 Universitetskaya nab., 199034 St. Petersburg, Russia}

\author{A.~V.~Stolyarov}
\affiliation{Department of Chemistry, Lomonosov Moscow State University, 1/3 Leninskie gory, 119991 Moscow, Russia}

\author{L.~V.~Skripnikov}
\affiliation{Department of Physics, St. Petersburg State University, 7/9 Universitetskaya nab., 199034 St. Petersburg, Russia}
\affiliation{National Research Centre “Kurchatov Institute” B.P. Konstantinov Petersburg Nuclear Physics Institute, 188300 Gatchina, Leningrad district , Russia}

\author{V.~M.~Shabaev}
\affiliation{Department of Physics, St. Petersburg State University, 7/9 Universitetskaya nab., 199034 St. Petersburg, Russia}
\affiliation{National Research Centre “Kurchatov Institute” B.P. Konstantinov Petersburg Nuclear Physics Institute, 188300 Gatchina, Leningrad district , Russia}

\author{I.~I.~Tupitsyn}
\affiliation{Department of Physics, St. Petersburg State University, 7/9 Universitetskaya nab., 199034 St. Petersburg, Russia}

\date{\today}

\begin{abstract}
 $\textit{Ab initio}$ calculations of the parallel component of the magnetic dipole hyperfine structure (HFS) constant have been carried out for hydroxyl radical isotopologues ($^{16,17}$OH(D)) over the internuclear distance range $R \in [0.6, 1.8]$ \AA.  For the ground electronic state $X^2\Pi$, the HFS functions were evaluated for contributions induced by both oxygen and hydrogen nuclei. In addition, the hydrogen-induced HFS curve was calculated for the excited $A^2\Sigma^+$ state. The quantum-chemistry study employs a four-component relativistic coupled-cluster (CC) method, including excitations up to the triple level, namely: the contribution of triple-cluster amplitudes was studied both perturbatively (CCSD(T)) and through fully iterative calculations (CCSDT). 
The resulting oxygen- and hydrogen-induced HFS functions represent the most accurate and reliable theoretical predictions to date exhibiting excellent agreement with semiempirical curve for hydrogen-induced HFS derived from high-resolution spectroscopic data for the lowest vibrational levels  ($v\in [0,2]$) of the electronic $X^2\Pi$ state. 
Vibrationally averaged $\textit{ab initio}$ values are consistent with experimental values within $1\%$ for all states considered. 
Furthermore, the internuclear distance range over which the HFS curves are defined has been extended beyond that of previous studies, thereby providing a robust foundation for accurate HFS treatments of higher-lying rovibrational levels of OH isotopologues within both adiabatic and non-adiabatic frameworks.
 
\end{abstract}

\maketitle
\section{Introduction}\label{sec:intro} 
The hydroxyl radical (OH) was among the first diatomic molecules to be identified in the Universe through astronomical radio spectroscopy. It exhibits a rich fine and hyperfine structure (HFS), characterized by spin–orbit coupling and $\Lambda$-doubling effects, which are essential for the unambiguous interpretation of its microwave spectra. In contrast to the fine structure, the HFS of OH depends sensitively on the nuclear spins of both oxygen and hydrogen atoms, making it a valuable probe for distinguishing between different OH isotopologues.

The OH isotopologues play a fundamental role in gas-phase chemistry and astrophysics, with relevance to interstellar and photospheric processes, atmospheric chemistry, and combustion phenomena. In atmospheric and climate studies, the OH radical is recognized as the most important oxidizing agent in the troposphere~\cite{stone2012tropospheric}. It contributes to the terrestrial nightglow~\cite{cosby2007oh} and plays a role in regulating the density of the ozone layer~\cite{lelieveld2004role}. In combustion chemistry, OH is a key intermediate species, significantly influencing reaction rates, flame temperatures, and the formation of pollutants~\cite{kathrotia2010study}. In astrophysics, the interstellar OH masers are among the most common and well-studied sources of coherent cosmic emission. The HFS of OH transitions provides a powerful diagnostic tool~\cite{Wright_2023} for probing magnetic fields, temperatures, masses and radiation environments in various astrophysical settings~\cite{ebisawa2015oh, cotten2012hydroxyl}. 

For open-shell systems such as the hydroxyl radical, accurate \textit{ab initio} calculations are essential to capture the complexity of electronic structure, including fine and hyperfine interactions. Previous theoretical studies have primarily employed non-relativistic methods, such as multireference configuration interaction and many-body perturbation theory (MBPT), which often provide a limited treatment of electron correlation effects. Moreover, most of these studies focus solely on the HFS parameters estimated near the equilibrium bond length or for the selected rovibrational levels, without addressing their explicit dependence on the internuclear distance $R$. This work addresses these limitations by employing relativistic four-component coupled-cluster (CC) theory, incorporating high-order excitations, to compute the magnetic HFS functions across a wide range of $R$ for both the ground $X^2\Pi$ and excited $A^2\Sigma^+$ states of ($^{16,17}$OH(D)) isotopologues. The derived \textit{ab initio} HFS functions enable straightforward extension of accurate HFS treatments to the higher-lying rovibrational levels of OH isotopologues in the framework of both conventional adiabatic approximation and advanced coupled-channel deperturbation analysis~\cite{Yurchenko2022}.
\par
The paper is organized as follows. Section~\ref{sec:methods} provides a detailed description of the methodological approaches, computational methods and parameters used in this study. Section~\ref{sec:results} presents the calculated HFS constants for both the ground and excited electronic states. Finally,  Section~\ref{sec:concl} summarizes the main findings and conclusions of the work.

\section{Theoretical approaches and methods}\label{sec:methods}
The calculations employ the relativistic four-component Dirac–Coulomb (DC) Hamiltonian, as implemented in the DIRAC~\cite{Saue:2020} software package and its associated extensions.
The DC Hamiltonian for an electronic system is given by:
\begin{equation}
\label{eq:HDC}
    H_{\text{DC}} = \Lambda^+ \left[\sum_i c \vec{\alpha}_i \cdot \vec{p}_i + \beta_i c^2 + V(\vec{r}_i) + \sum_{i < j} \frac{1}{ |\vec{r}_i - \vec{r}_j|} \right] \Lambda^+,
\end{equation}
where $ \beta $ and $ \vec{\alpha} $ are the standard $ 4 \times 4 $ Dirac matrices, $\vec{r}_i$ and $\vec{p}_i$ denote the position and momentum operators of the $i$-th electron, $ V(\vec{r}_i) $ represents the total nuclear binding potential. We use atomic
units unless stated otherwise. The finite nuclear size is accounted for using a Gaussian model of charge distribution. The summation goes over all electrons in the system, and $ \Lambda^+ $ denotes the projection operator onto the positive-energy solutions of the Dirac-Fock one-electron equation.

In magnetic dipole approximation the relativistic HFS operator is expressed as the scalar product of the electronic operator $\boldsymbol{T}$ and the nuclear magnetic moment operator $\boldsymbol{\mu}$:
\[ H_\mathrm{hfs} = \boldsymbol{\mu} \cdot \boldsymbol{T}, \]
where
\[ \boldsymbol{T} = \sum_i \frac{[\mathbf{r}_i \times \boldsymbol{\alpha}_i]}{r_i^3}
\label{method:T} \]
and
\[ \boldsymbol{\mu} = \frac{\mu}{I} \boldsymbol{I} . \]
Here, $\mu$ is the nuclear magnetic moment, and  $\boldsymbol{I}$ is the nuclear spin operator. We shall throughout this paper give magnetic moments in units of the nuclear magneton $\mu_\mathrm{N}$. \par
We consider the "parallel"\ component of the HFS constant, defined as the diagonal matrix element of $H_\mathrm{hfs}$ for a given electronic state $\psi_n$ with total electronic angular momentum projection $\Omega$ onto the internuclear axis $z$:
\[A_{||} = \frac{\mu}{\Omega I}\braket{\psi_n|T_z|\psi_n}. \label{method:constant_def}\]
The diagonal matrix element in Eq.~(\ref{method:constant_def}) can be evaluated using the finite-field approach, in which the operator $T_z$ is added to the DC Hamiltonian $H_{\text{DC}}$ with a  scaling parameter $\lambda$. The expectation value is then obtained from the derivative of the total energy with respect to $\lambda$:
\[\braket{\psi_n|T_z|\psi_n} = \left. \frac{d E_n}{d \lambda}  \right|_{\lambda =0} \approx \frac{E_n(+\Delta\lambda)-E_n(-\Delta\lambda)}{2\Delta\lambda},\]
where a central finite-difference scheme is used to approximate the derivative numerically.
The choice of $\Delta\lambda$ is crucial for ensuring the numerical stability and accuracy of the results. In this work, $\Delta\lambda$ is set to $1.3 \cdot 10^{-4}$ a.u..

\par 
In the present work, the HFS constants have been calculated for several isotopologues of the hydroxyl radical. For the ground electronic doublet state X$^2\Pi$, calculations were performed for the HFS induced by the hydrogen nucleus ($^1$H, $I = 1/2$, $\mu = 2.79284 \mu_\mathrm{N}$) in the  $^{16}\mathrm{O}\mathrm{H}$ molecule and by the oxygen nucleus ($^{17}$O, $I = 5/2$, $\mu = -1.89379 \mu_\mathrm{N}$) in the $^{17}\mathrm{O}\mathrm{H}$ molecule.  
For the excited A$^2\Sigma^+$ state of $^{16}\mathrm{O}\mathrm{H}$, the hydrogen-induced HFS was investigated. Additionally, the HFS induced by the deuterium nucleus ($^2$D, $I = 1$, $\mu = 0.85743 \mu_\mathrm{N}$) was calculated for the A$^2\Sigma^+$ state of the $^{16}\mathrm{O}\mathrm{D}$ molecule. The nuclear magnetic moment values were taken from~\cite{sansonetti2005}. All HFS constants were calculated as functions of the internuclear distance over the range $0.6 \le R \le 1.8$ \AA.
To enable comparison with experimental data, the obtained HFS constants were averaged over vibrational wave functions derived from semiempirical potential energy curves. For the ground state, a highly accurate semiempirical potential energy curve from Ref.~\cite{augustovivcova2020} was used, whereas the relevant semi-empirical potential for the excited state was derived using a direct-potential-fit of the experimental A-X spectra of both OH and OD isotopologues, as reported in Ref.~\cite{Pazyuk2025}. Although  Ref.~\cite{augustovivcova2020} provides data for the OH molecule with oxygen isotopes $^{16}$O and $^{18}$O, it does not include $^{17}$O. However, our calculations indicate that the contribution from the mass-dependent part of the potential to the $^{17}$O-induced HFS constant is negligible, less than $0.001$~MHz for the lowest vibrational level, and thus below the theoretical uncertainty.
\par
In most of the studies, the HFS constants of diatomic molecules are reported in terms of the Frosch and Foley parameters $a,b_\mathrm{F},c,d$. These parameters define the matrix elements of effective operators that arise from the reduction of the four-component Dirac equation to the two-component Pauli form~\cite{frosch1952}. The relations between the $\textit{ab initio}$ HFS constant $A_{||}$ and the Frosch and Foley parameters can be formulated within different Hund's coupling cases, as demonstrated, for example, in Ref.~\cite{malika2022}. \par
For the ground X$^2\Pi$ state of the OH molecule, which corresponds to Hund's case (a) the hyperfine constant $A_{||}$ is related to the Frosch and Foley parameters by the following expressions:
\[A_{||}\left( ^2\Pi_\frac{1}{2} \right) = 2 a - b_\mathrm{F} - \frac{2}{3}c, \]
 \[A_{||}\left( ^2\Pi_\frac{3}{2} \right) = \frac{2}{3} a + \frac{1}{3}b_\mathrm{F} + \frac{2}{9}c. \]
For the excited A$^2\Sigma^+$ state, which is better described by Hund’s case (b), the relationship is:
 \[A_{||}\left( ^2\Sigma^+ \right) =b_\mathrm{F} + \frac{2}{3}c.\]
These expressions are employed to facilitate comparison with other experimental and theoretical results reported in terms of the Frosch and Foley parameters.
\par 
The four-component relativistic single-reference
coupled-cluster method was employed, incorporating single, double, and perturbative triple excitations (CCSD(T)), as implemented in the EXP-T package~\cite{oleynichenko2020towards, Oleynichenko:website}.
The hyperfine interaction operator given in Eq.~(\ref{method:T}) introduced at the coupled-cluster level, following the construction of the one-electron basis using the~{Dirac-Hartree-Fock} method in the DIRAC program. CC calculations were performed with correlation of all $9$ electrons in the OH and OD molecules, and virtual orbitals were included up to an energy cut-off of $5000$~a.u.. Core-valence  correlation-consistent basis sets aug-cc-pCV$n$Z ($n=4,5$) were  employed, enabling extrapolation of the total energy to the complete basis set (CBS) limit using the following formula: 
\[ E(n) = E^\mathrm{CBS} + \frac{A}{n^3}. \label{methods:CBS}\]


\par
The extrapolation formula~(\ref{methods:CBS}) was additionally validated, and its associated error was estimated. To this end, basis sets for hydrogen and oxygen were optimized by calculations the hydrogen-induced HFS constant by the CCSD method at the equilibrium internuclear distance ($R_e = 0.96966$ \AA~\cite{huber2013}) of the X$^2\Pi_{3/2}$ state of the  $^{16}\mathrm{O}\mathrm{H}$ molecule. The optimization procedure followed a strategy similar to that described in Refs~\cite{mosyagin2000comparison, kaygorodov2021electron, kaygorodov2022ionization}.  The optimization began with the dyall.ae3z set of primitive Gaussian functions, comprising $9s2p1d$ functions for hydrogen and $14s8p3d1f$ for oxygen. Additional basis functions were incrementally added to this set, with exponents chosen to maximize the contribution of each function to the HFS constant. The final optimized basis set included additional $6s6p4d4f1g1h$ functions for hydrogen and $1s4p4d2f1g1h$ functions for oxygen. To evaluate the  residual error due to basis set incompleteness, Table~\ref{Tables:optimization} presents the variation of the HFS constant upon adding each type of function to the optimized basis set. The squared  sum of these variations yields a total estimated uncertainty of $0.04$~MHz, corresponding to less than $0.1\%$ of the final CCSD value of $63.950$~MHz at the equilibrium distance.

\setlength{\tabcolsep}{6pt}
\begin{table}[H]
\centering
\caption{Variation of hydrogen-induced HFS constant at $R_e$ due to the inclusion of additional basis functions beyond the optimized basis set at the CCSD level.}
\begin{tabular}{c||ccccccc}
                              & $s$           & $p$            & $d$  & $f$  & $g$  &$h$ & $i$  \\ \hline
H, $10^{-3}$ MHz & $1$ & $3$            & $7$ & $4$  & $12$ & $5$ & $15$ \\
O, $10^{-3}$ MHz & $3$ & $1$ & $3$  & $13$ &  $24$  & $8$ &  $20$
\end{tabular}
\label{Tables:optimization}
\end{table}
  The final CCSD(T) values for the hydrogen-induced HFS function were obtained using the  optimized basis set. The associated basis set error was estimated as the difference between the values calculated with the optimized basis set and those obtained via CBS extrapolation using Eq.~(\ref{methods:CBS}). A more detailed comparison of these results is presented in the following section. \par 
 For the oxygen-induced HFS function, the CCSD(T) values were determined through the CBS extrapolation~(\ref{methods:CBS}) as well. The basis set error in this case was estimated as the difference between the CBS-extrapolated value and the result obtained using the aug-cc-pCV$5$Z basis set. 
 \par
 
 Additionally, the contribution of full triple excitations in the CC method was investigated. This contribution was determined as the difference between the CCSDT and CCSD(T) results. For the X$^2\Pi$ state, calculations were performed using the cc-pCVQZ basis set, while for the A$^2\Sigma^+$ state, the dyall.ae4z basis set was employed. The difference between the CCSDT and CCSD(T) results was also used to estimate the uncertainty associated with the incomplete treatment of higher-order excitations.

\section{Results and Discussion}\label{sec:results}

\subsection{The ground X$^2\Pi_{1/2;3/2}$ state}\label{sec:results_ground}
\subsubsection{$\text{H-induced HFS}$}\label{sec:results_ground}
\begin{figure}[H]
\centering
\includegraphics[width=1\linewidth,clip]{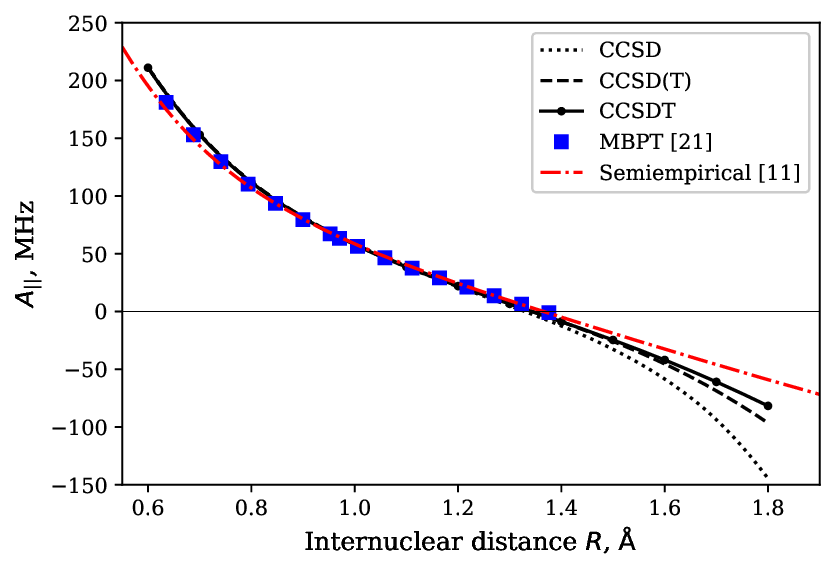}
\caption{\label{Fig:1H_X2Pi32}
Calculated $\mathrm{H}$-induced HFS function $A_{||}$ for X$^2\Pi_{3/2}$ state of the $^{16}$OH molecule. The semiempirical curve taken from Augustovi{\v{c}}ov{\'a} and {\v{S}}pirko~\cite{augustovivcova2020}, as well as MBPT data from Kristiansen and Veseth~\cite{kristiansen1986part1}, are also shown.}
\end{figure}

The hydrogen-induced HFS function for the ground X$^2\Pi_{3/2}$ state of $^{16}$OH molecule is represented in Figure~\ref{Fig:1H_X2Pi32}. It has a strong dependence on the internuclear distance, varying from $200$~MHz at $0.6$~\AA \ to $-80$~MHz at $1.8$~\AA, passing through zero around $1.4$~\AA. This behavior can be attributed to a strong change in the electronic structure, in particular, the electron charge on the hydrogen atom at different internuclear distances.
\par 
A significant contribution from triple excitations in the CCSD(T) and CCSDT methods, relative to CCSD, emerges at interatomic distances exceeding $1.4$~\AA. In this region, the maximum coupled-cluster amplitudes reach approximately $0.1$ and $0.3$ for single and double excitations, respectively. Consequently, the inclusion of higher-order excitations is necessary for achieving more accurate results. At the largest available distance of $1.8$~\AA, the perturbative triple excitations contribute up to $80$~MHz, with an additional $20$~MHz arising from the inclusion of full triples.\par
For the HFS induced by the hydrogen nucleus, semiempirical functions are available from Ref.~\cite{augustovivcova2020}. It is important to note that the semiempirical curve is based on experimental data for vibrational levels $v \le 2$, corresponding to internuclear distances in the range of $0.8$ to $1.3$~\AA, and on calculations from Ref.~\cite{kristiansen1986part1}, which employed many-body perturbation theory (MBPT). These theoretical results are also shown in the figure. For internuclear distances beyond $1.4$~\AA, the semiempirical data rely on an extrapolation of both experimental and theoretical data.
As shown in Figure~\ref{Fig:1H_X2Pi32}, excellent agreement is observed between our results and the reference data in the region up to $1.4$~\AA. At larger distances, however, the CC curves exhibit a more pronounced decrease than predicted by the semiempirical extrapolation.\par

\begin{figure}[H]
\centering
\includegraphics[width=1\linewidth,clip]{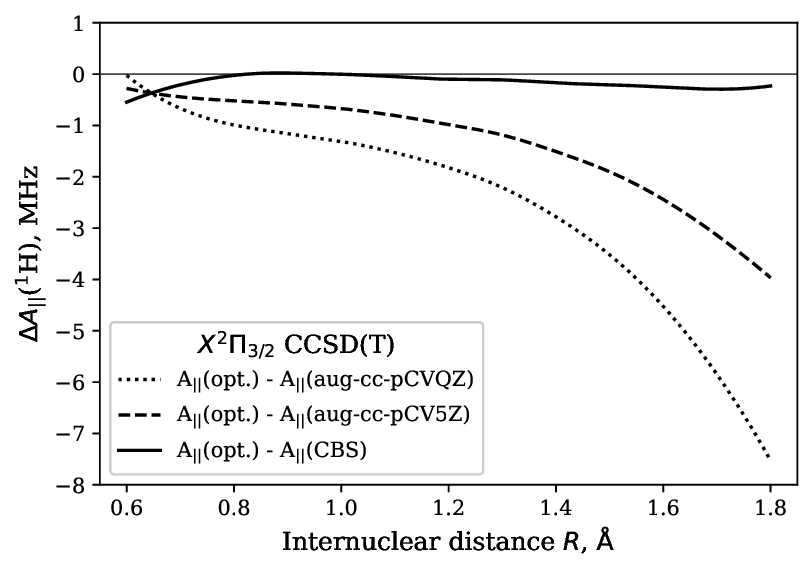}
\caption{\label{Fig:CBSvsOPT}
Basis set convergence of the $\mathrm{H}$-induced HFS constant $A_{||}$ for the X$^2\Pi_{3/2}$ state. The results are based on the aug-cc-pCV$n$Z basis set family and are shown relative to those obtained with the optimized basis set within the CCSD(T) approximation.
}
\end{figure}

 To assess the consistency between results obtained using the optimized basis set and those derived from extrapolation~(\ref{methods:CBS}) with standard aug-cc-pCV$n$Z ($n=4,5$) basis sets, a comparison of the two approaches is presented in Fig.~\ref{Fig:CBSvsOPT}. The values calculated with the optimized basis set are taken as the reference (zero baseline). The figure demonstrates that the extrapolated values closely reproduce those obtained with the optimized basis set, indicating high accuracy of the extrapolation procedure. Notably, although the basis set optimization was performed at the equilibrium bond length, the agreement persists across the entire range of internuclear distances considered.

\begin{figure}[H]
\centering
\includegraphics[width=1\linewidth,clip]{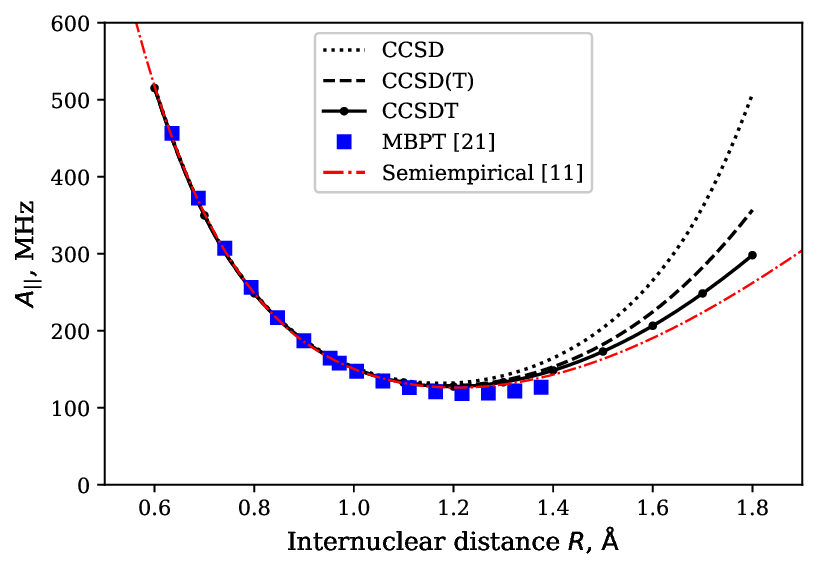}
\caption{\label{Fig:1H_X2Pi12}
Calculated $\mathrm{H}$-induced HFS function $A_{||}$ for X$^2\Pi_{1/2}$ state of the $^{16}$OH molecule. The semiempirical curve taken from Augustovi{\v{c}}ov{\'a} and {\v{S}}pirko~\cite{augustovivcova2020}, as well as MBPT data from Kristiansen and Veseth~\cite{kristiansen1986part1}, are also shown.
}
\end{figure}

For the other component of the ground-state doublet, X$^2\Pi_{1/2}$, the hydrogen-induced HFS function exhibits the behavior shown in Fig.~\ref{Fig:1H_X2Pi12}. In contrast to the X$^2\Pi_{3/2}$ state, the HFS function does not change sign; instead, the curve displays a flat minimum near $1.2$~\AA. Beyond this minimum, the inclusion of higher-order cluster amplitudes becomes increasingly important. At an internuclear distance of $1.8$~\AA, the contribution from perturbative triple excitations reaches approximately $150$~MHz, with an additional $60$~MHz arising from the inclusion of full triples. As with the X$^2\Pi_{3/2}$ state, our final results are in good agreement with the semiempirical curve up to $1.4$~\AA, beyond which noticeable deviations begin to appear.

To facilitate comparison with the experimental data, the values of the hydrogen-induced HFS functions averaged over the vibrational wave functions $v \le 2$ are given in Table~\ref{Tables:1H-HFS_X2Pi}. Since the internuclear distances relevant to these vibrational states lie within the region where the results from both the CBS extrapolation and the optimized basis set are in excellent agreement (see Fig.~\ref{Fig:CBSvsOPT}), the dominant source of uncertainty in our calculations arises from the incomplete treatment of higher-order excitations in the CC method. The contribution from triple excitations increases with vibrational quantum number. Perturbative inclusion of triple excitations via the CCSD(T) method contributes approximately $0.5-1.5 \%$ to the final values for the vibrational levels considered, while the inclusion of full triples using the CCSDT method adds a further $0.2-0.5 \%$.
The final computed values show excellent agreement with the experimental data, with deviations below $1\%$. 
Theoretical results, obtained using MBPT~\cite{kristiansen1986part1} and CASSCF/MRCI method~\cite{chong1991theoretical}, show a similar level of agreement with the experiment.

\subsubsection{$^{17}\text{O-induced HFS}$}\label{sec:results_ground_oxygen}
\begin{figure}[H]
\centering
\includegraphics[width=1\linewidth,clip]{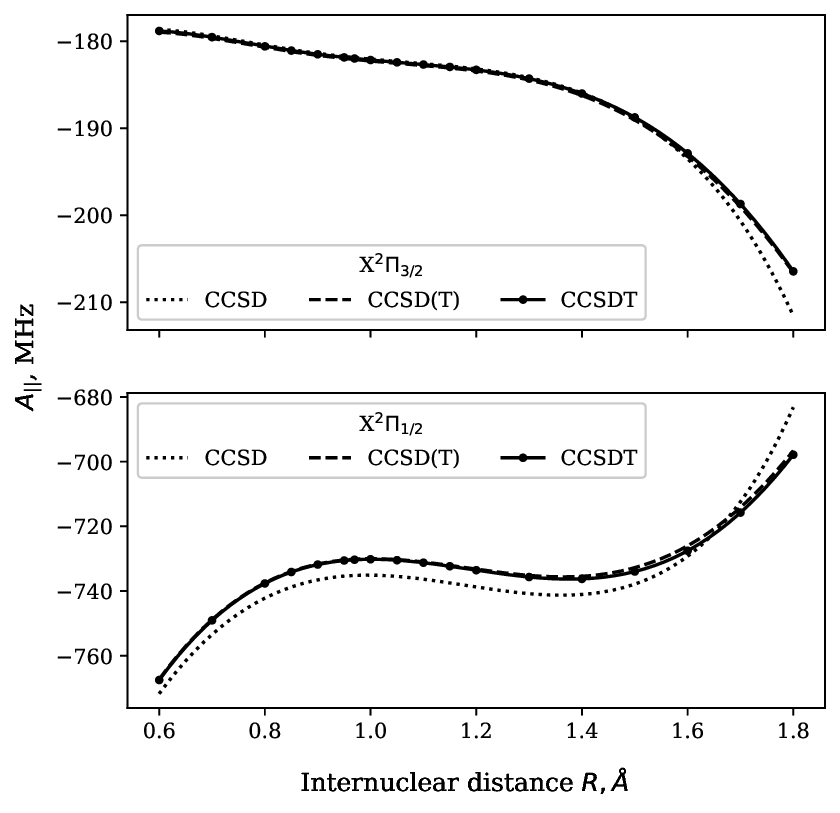}
\caption{\label{Fig:17O_X2Pi}
Calculated $^{17}\mathrm{O}$-induced HFS function $A_{||}$ for the X$^2\Pi_{1/2}$ and X$^2\Pi_{3/2}$ states of the $^{17}$OH molecule.
}
\end{figure}

The results for oxygen-induced HFS function in the ground state are presented in Figure~\ref{Fig:17O_X2Pi}. In contrast to the hydrogen-induced case, the curves in Figure~\ref{Fig:17O_X2Pi} exhibit relatively weak dependence on the internuclear distance. In the non-relativistic limit this behavior can be explained by the negligible contribution of the Fermi contact term. Specifically, the unpaired electron in the $\pi$ orbital does not significantly polarize the inner $1s$ and $2s$ orbitals of the oxygen atom. As a result, the HFS is predominantly determined by the spin-dipole and spin-orbit terms, which exhibit a much weaker dependence on bond length.

\par 
For the X$^2\Pi_{3/2}$ state, perturbative triple excitations contribute approximately $0.5-0.8\%$ relative to the CCSD values up to $1.4$~\AA, after which the contribution increases sharply, reaching nearly $2\%$ at $1.8$~\AA. In the case of the X$^2\Pi_{1/2}$ state, the contribution is around $0.1$-$0.2\%$ up to $1.4$~\AA \ -- but similarly rises to several percent at larger internuclear distances. Inclusion of full triple excitations via the CCSDT method, in comparison with the perturbative triples in CCSD(T), does not result in a significant change in the shape or behavior of the curves.
\par

The values of the oxygen-induced HFS function, averaged over the vibrational wave functions, are summarized in Table~\ref{Tables:17O-HFS_X2Pi}. It is evident that the inclusion of excitations beyond CCSD does not significantly affect the results for these states. Our computed values for the lowest vibrational level show good agreement with the experimental data reported in Ref.~\cite{leopold1987, drouin2013isotopic} for both components of the ground-state doublet. In contrast, the available theoretical results obtained using CASSCF/MRCI approach~\cite{chong1991theoretical} exhibit noticeably larger deviations from experiment. Consequently, the reported dependence of the HFS function on internuclear distance in Ref.~\cite{chong1991theoretical} appears unreliable, and we therefore refrain from a detailed comparative analysis of its distance dependence.

\subsection{The excited A$^2\Sigma^+$ state}\label{sec:results_excited}
\begin{figure}[H]
\centering
\includegraphics[width=1\linewidth,clip]{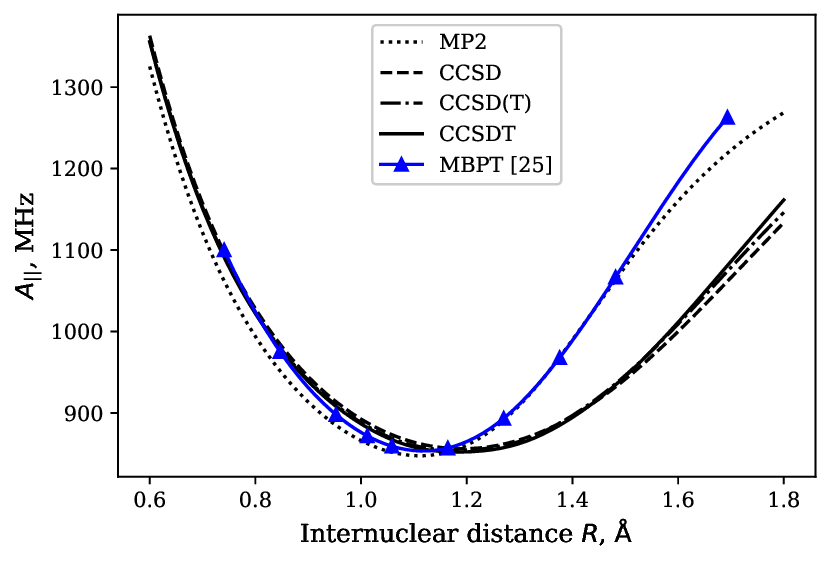}
\caption{\label{Fig:1H_A2Sigma}
 Calculated $\mathrm{H}$-induced HFS function $A_{||}$ for the A$^2\Sigma^+$ state  of the $^{16}$OH molecule.  Theoretical MBPT data from Kristiansen and Veseth~\cite{kristiansen1986part2} are also displayed.
}
\end{figure}

The H- and D-induced HFS functions of $^{16}$OH and $^{16}$OD molecules were calculated for the first excited A$^2\Sigma^+$ state.  Since both nuclei involve the same electronic structure, Figure~\ref{Fig:1H_A2Sigma} shows the HFS constant curve for hydrogen only. The corresponding curve for the deuterated hydroxyl radical can be obtained by scaling the hydrogen curve by a factor $\mu_\mathrm{D}/(2\mu_\mathrm{H}) \approx 0.1535$. In the non-relativistic approach, the spin-orbit contribution vanishes, and the Fermi contact interaction becomes the dominant term determining the HFS function. \par

As shown in Fig.~\ref{Fig:1H_A2Sigma}, the available theoretical curve obtained using third-order MBPT~\cite{kristiansen1986part1} deviates significantly from our results computed using the coupled-cluster method. To further illustrate this discrepancy, the figure also includes results from second-order Møller–Plesset perturbation theory (MP2), which exhibits similar behavior to the MBPT curve. This comparison highlights the limitations of such perturbative approaches in accurately describing the HFS in contrast to the coupled-cluster method. \par 
Table~\ref{Tables:HFS_A2Sigma} presents the HFS constants of the $^{16}$OH and $^{16}$OD molecules,  averaged over the lowest vibrational states. The perturbative contribution of triple excitations within the coupled-cluster framework accounts for up to $0.5\%$ of the final values, while the additional inclusion of full triple excitations contributes approximately $0.2\%$ across all states considered. For the $^{16}$OH molecule, our final result for the $v=0$ state shows excellent agreement with all available experimental data~\cite{raab1981precision,ter1983determination,ter1986observation,fast2018precision}. However, while experimental measurements~\cite{ter1986observation} report a decrease of approximately 10 MHz in the HFS constant for the $v = 1$ state, such a reduction is not observed in our calculations. In contrast, theoretical predictions from Ref.~\cite{kristiansen1986part2} indicate an increase in the HFS constant with vibrational excitation.
For the $^{16}$OD molecule, the vibrational dependence of the averaged HFS constant is significantly weaker, in full agreement with experimental findings~\cite{german1976high, xin2003energy}. Other available theoretical results based on CI methods~\cite{green1973dipole} exhibit larger discrepancies from experiment, particularly showing a more pronounced variation between vibrational states.

\par
\begin{figure}[H]
\centering
\includegraphics[width=1\linewidth,clip]{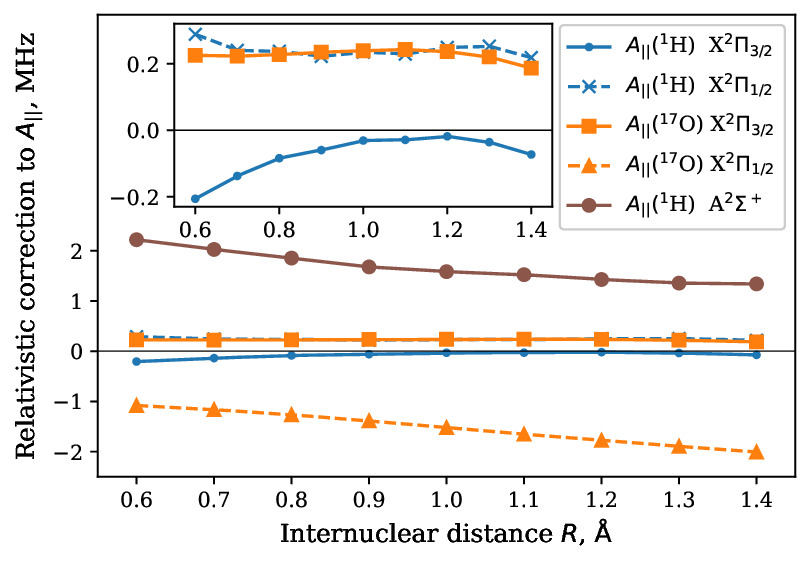}
\caption{\label{Fig:REL_CORR}
 Relativistic correction to the HFS function $A_{||}$ for the X$^2\Pi_{1/2, 3/2}$ and A$^2\Sigma^+$ states of the OH molecule for both $\mathrm{H}$- and 
$^{17}\mathrm{O}$-induced hyperfine interactions.
}
\end{figure}

To assess the influence of relativistic effects on the HFS functions, we calculated 
the $A_{||}$  values using both the standard speed of light ($c=137.036$~a.u.) and an approximate nonrelativistic limit ($c\xrightarrow{} \infty$), the latter modeled by increasing the standard $c$ value by a factor of $50$. Fig.~\ref{Fig:REL_CORR} presents the resulting relativistic corrections to the HFS functions for all investigated states, obtained with the CCSD method and the aug-cc-pCVQZ basis set. As shown, relativistic effects are minor. For the $\mathrm{H}$-induced HFS of the X$^2\Pi_{3/2}$ state and the $^{17}\mathrm{O}$-induced HFS of the X$^2\Pi_{1/2}$ state, the correction decreases the $A_{||}$
value, whereas in all other cases it leads to a slight increase.

\section{Concluding remarks}\label{sec:concl}
This work presents \textit{ab initio} investigation of the magnetic dipole hyperfine structure (HFS) functions of the hydroxyl radical using relativistic four-component coupled-cluster theory. The study was carried out for the underlying electronic X$^2\Pi$ and A$^2\Sigma^+$ states considering HFS contributions induced by the hydrogen nucleus in both states and by the oxygen nucleus in the ground state.

The sensitivity of the results to the choice of basic sets and the level treatment of coupled-cluster excitations, was systematically investigated to ensure high accuracy and to estimate associated uncertainties. In the case of hydrogen-induced HFS, a detailed comparison was made between results obtained using a specially optimized basis set and those derived from extrapolation formulas applied to standard correlation-consistent basis sets. To evaluate the impact of higher-order excitations beyond the CCSD level, calculations were performed  using both the perturbative CCSD(T) and the fully iterative CCSDT methods. The contribution of triple cluster amplitudes becomes significant at large internuclear distances $R > 1.4$~\AA. However, for the lower vibrational levels where experimental data are available, this contribution is relatively small, ranging from $0.5\%$ to $2\%$.
 
The HFS functions averaged over vibrational wave functions are consistent with experimental values on the order of $1\%$ across the both states considered. The dependence of the HFS functions on the internuclear distance was systematically examined, revealing significant variation for the hydrogen-induced HFS, in contrast to the more monotonic behavior observed for the oxygen-induced HFS. Notably, the computed hydrogen-induced HFS function for the ground state coincides with high accuracy with the available semiempirical curve within the range of distances relevant to experimental measurements. 

This work demonstrates the effectiveness of relativistic quantum-chemical methods in accurately predicting HFS in molecular systems. The derived \textit{ab initio} HFS functions provide more accurate and reliable results for the oxygen- and hydrogen-induced HFS in both the ground and excited electronic states of the OH molecule. 
Additionally, the range of internuclear distances for which the HFS curves are defined has been extended beyond that considered in previous studies. Consequently, the resulting functions can be readily applied for the accurate treatment of HFS in higher-lying rovibrational levels of OH isotopologues within both adiabatic and non-adiabatic frameworks. \\

\subsection*{\bf Acknowledgments}
The study was supported by the Russian Science Foundation (interdisciplinary grant No. 22-62-00004). Computations were held on the basis of the HybriLIT heterogeneous computing platform (LIT, JINR).
%
%
\onecolumngrid\
\begin{table}[H]
\centering
    \caption{$A_{||}$ constant (MHz) for the HFS, induced by the H nucleus for the X$^2\Pi$ states of the $^{16}$OH molecule. The HFS constants are averaged over the vibrational wave functions.}
    
\begin{tabular}{lc|S[table-format=3.5]S[table-format=3.5]S[table-format=3.6]|S[table-format=2.5]S[table-format=2.5]S[table-format=2.5]}
\multicolumn{1}{c}{}       & 
\multicolumn{1}{c|} {~~~Ref.~~~}  &
\multicolumn{3}{c|}{X$^2\Pi_{1/2}$} & \multicolumn{3}{c}{X$^2\Pi_{3/2}$} \\ 
\hline
 &
 &\multicolumn{1}{l}{$v=0$}           & \multicolumn{1}{l}{$v=1$}          & \multicolumn{1}{l|}{$v=2$}          & \multicolumn{1}{l}{$v=0$}          & \multicolumn{1}{l}{$v=1$}          & \multicolumn{1}{l}{$v=2$}          \\ \hline
CCSD &    & 160.9           & 161.0          & 162.4          & 61.2           & 55.8           & 50.1           \\
CCSD(T) &  & 159.5           & 159.2          & 160.0          & 61.6           & 56.3           & 50.8           \\
CCSDT &  & 159.2(4)        & 158.7(5)       & 159.3(7)       & 61.7(1)        & 56.5(2)        & 51.0(3)         \\[1ex]
MBPT & \cite{kristiansen1986part1} & 157.41          & 155.94         & 154.83         & 61.26          & 56.41          & 51.03          \\
CASSCF/MRCI &  \cite{chong1991theoretical}            & 157.23          & 157.26         &                & 61.71          & 55.59          &                \\[1ex]
Expt. & \cite{coxon1980}                            & 158.282(14)     & 157.694(35)    & 158.081(41)    & 62.054(5)      & 56.866(12)     & 51.480(14) 
\end{tabular}

    \label{Tables:1H-HFS_X2Pi}
\end{table}
\begin{table}[H]
\centering
    \caption{$A_{||}$ constant (MHz) for the HFS, induced by the $^{17}$O nucleus for the X$^2\Pi$ states of the $^{17}$OH molecule. The HFS constants are averaged over the vibrational wave functions.}
    \begin{tabular}{lc|S[table-format=4.6]S[table-format=4.4]|S[table-format=4.6]S[table-format=4.4]}
                \multicolumn{1}{c}{}       & 
\multicolumn{1}{c|} {~~~Ref.~~~}     & \multicolumn{2}{c|}{X$^2\Pi_{1/2}$} & \multicolumn{2}{c}{X$^2\Pi_{3/2}$} \\ \hline

               &     & \multicolumn{1}{l}{$v=0$}               & \multicolumn{1}{l|}{$v=1$}                           & \multicolumn{1}{l}{$v=0$}             & \multicolumn{1}{l}{$v=1$}                          \\ \hline
CCSD      &      & -736               & -737                           & -181.9             & -182.1                        \\
CCSD(T)    &     & -731               & -732                           & -182.2             & -182.3                        \\
CCSDT     &      & -731(3)            & -732(3)                        & -182.3(6)          & -182.4(6)                     \\ [1ex]
CASSCF/MRCI & \cite{chong1991theoretical}    & -716.37            & -715.62                        & -186.63            & -187.51                    \\  [1ex]
Expt. & \cite{leopold1987} & -737.4(34)         &          & -181.8(11)         &    \\
Expt. & \cite{drouin2013isotopic} & -729.816(138)         &          & -182.106(46)         &         

\end{tabular}
    \label{Tables:17O-HFS_X2Pi}
\end{table}
\twocolumngrid\
\onecolumngrid\
\begin{table}[H]
\centering
    \caption{$A_{||}$ constant (MHz) for the HFS, induced by the H(D) nucleus for the A$^2\Sigma^+$ state of the $^{16}$OH ($^{16}$OD) molecule. The HFS constants are averaged over the vibrational wave functions.}
    \begin{tabular}{lc|S[table-format=4.5]S[table-format=4.4]|S[table-format=3.4]S[table-format=3.4]S[table-format=3.4]S[table-format=3.4]}
                        \multicolumn{1}{c}{}       & 
\multicolumn{1}{c|} {~~~Ref.~~~}     & \multicolumn{2}{c|}{OH}     & \multicolumn{4}{c}{OD}                                                              \\ \hline
                         &   & \multicolumn{1}{l}{$v=0$}                & \multicolumn{1}{l|}{$v=1$}     & \multicolumn{1}{l}{$v=0$}                & \multicolumn{1}{l}{$v=1$}                & \multicolumn{1}{l}{$v=2$}      & \multicolumn{1}{l}{$v=3$}      \\ \hline 
CCSD               &         & 887                  & 888       & 136.2                & 136.3                & 137.0     & 138.4     \\
CCSD(T)      &               & 883                  & 884       & 135.6               & 135.6               & 135.7     & 136.5     \\
CCSDT         &              & 881(5)               & 882(5)    & 135.3(9)             & 135.3(9)             & 135.6(9)   & 136.3(9)   \\ [1ex]
MBPT & \cite{kristiansen1986part2} & 875.7                & 886.0     &  &  &                                &                                \\
CI & \cite{green1973dipole}                    &  &                                & 129.1                & 140.0                &                                &                                \\ [1ex]
Expt. & \cite{german1976high}                &  &                                & 135.46(29)           & 135.55(29)           &                                &                                \\
Expt. & \cite{xin2003energy}            &  &                                & 135.03(18)           & 135.10(26)           & 135.41(42) & 135.89(42) \\
Expt. &  \cite{raab1981precision}             & 886.7(9)             &                                &  &  &                                &                                \\
Expt.& \cite{ter1983determination} & 886.0(30)            & 875.8(26) &  &  &                                &                                \\
Expt. & \cite{ter1986observation} & 879.1(6)             &                                &  &  &                                &                                \\
Expt. & \cite{fast2018precision}             & 879.898(73)          &                                & 135.04(8)            &  &                                &                               
\end{tabular}

    \label{Tables:HFS_A2Sigma}
\end{table}
\twocolumngrid\
\newpage

\bibliographystyle{my-h-physrev.bst}
\bibliography{bibliography}

\begin{thebibliography}{10}

\bibitem{stone2012tropospheric}
D.{~}Stone, L.~K.{~}Whalley, and D.~E.{~}Heard,
\newblock Chemical Society Reviews {\bf 41},~6348 (2012).

\bibitem{cosby2007oh}
P.{~}Cosby and T.{~}Slanger,
\newblock Canadian journal of physics {\bf 85},~77 (2007).

\bibitem{lelieveld2004role}
J.{~}Lelieveld, F.{~}Dentener, W.{~}Peters, and M.{~}Krol,
\newblock Atmospheric Chemistry and Physics {\bf 4},~2337 (2004).

\bibitem{kathrotia2010study}
T.{~}Kathrotia, M.{~}Fikri, M.{~}Bozkurt, M.{~}Hartmann, U.{~}Riedel, and C.{~}Schulz,
\newblock Combustion and Flame {\bf 157},~1261 (2010).

\bibitem{Wright_2023}
S.~O.~M.{~}Wright, S.~K.{~}Nugroho, M.{~}Brogi, N.~P.{~}Gibson, E.~J.~W.{~}de~Mooij, I.{~}Waldmann, J.{~}Tennyson, H.{~}Kawahara, M.{~}Kuzuhara, T.{~}Hirano, T.{~}Kotani, Y.{~}Kawashima, K.{~}Masuda, J.~L.{~}Birkby, C.~A.{~}Watson, M.{~}Tamura, K.{~}Zwintz, H.{~}Harakawa, T.{~}Kudo, K.{~}Hodapp, S.{~}Jacobson, M.{~}Konishi, T.{~}Kurokawa, J.{~}Nishikawa, M.{~}Omiya, T.{~}Serizawa, A.{~}Ueda, S.{~}Vievard, and S.~N.{~}Yurchenko,
\newblock The Astronomical Journal {\bf 166},~41 (2023).

\bibitem{ebisawa2015oh}
Y.{~}Ebisawa, H.{~}Inokuma, N.{~}Sakai, K.~M.{~}Menten, H.{~}Maezawa, and S.{~}Yamamoto,
\newblock The Astrophysical Journal {\bf 815},~13 (2015).

\bibitem{cotten2012hydroxyl}
D.~L.{~}Cotten, L.{~}Magnani, E.~A.{~}Wennerstrom, K.~A.{~}Douglas, and J.~S.{~}Onello,
\newblock The Astronomical Journal {\bf 144},~163 (2012).

\bibitem{Yurchenko2022}
Q.{~}Qu, S.~N.{~}Yurchenko, and J.{~}Tennyson,
\newblock Journal of Chemical Theory and Computation {\bf 18},~1808 (2022), https://doi.org/10.1021/acs.jctc.1c01244,
\newblock PMID: 35148098.

\bibitem{Saue:2020}
T.{~}Saue, R.{~}Bast, A.~S.~P.{~}Gomes, H.~J.~A.{~}Jensen, L.{~}Visscher, I.~A.{~}Aucar, R.{~}Di~Remigio, K.~G.{~}Dyall, E.{~}Eliav, E.{~}Fasshauer, T.{~}Fleig, L.{~}Halbert, E.~D.{~}Hedegård, B.{~}Helmich-Paris, M.{~}Iliaš, C.~R.{~}Jacob, S.{~}Knecht, J.~K.{~}Laerdahl, M.~L.{~}Vidal, M.~K.{~}Nayak, M.{~}Olejniczak, J.~M.~H.{~}Olsen, M.{~}Pernpointner, B.{~}Senjean, A.{~}Shee, A.{~}Sunaga, and J.~N.~P.{~}van Stralen,
\newblock J. Chem. Phys. {\bf 152},~204104 (2020).

\bibitem{sansonetti2005}
J.~E.{~}Sansonetti and W.~C.{~}Martin,
\newblock Journal of physical and chemical reference data {\bf 34},~1559 (2005).

\bibitem{augustovivcova2020}
L.~D.{~}Augustovi{\v{c}}ov{\'a} and V.{~}{\v{S}}pirko,
\newblock Journal of Quantitative Spectroscopy and Radiative Transfer {\bf 254},~107211 (2020).

\bibitem{Pazyuk2025}
E.~A.{~}Pazyuk and A.~V.{~}Stolyarov,
\newblock private communication, 2025.

\bibitem{frosch1952}
R.~A.{~}Frosch and H.{~}Foley,
\newblock Physical Review {\bf 88},~1337 (1952).

\bibitem{malika2022}
M.{~}Denis, P.~A.{~}Haase, M.~C.{~}Mooij, Y.{~}Chamorro, P.{~}Aggarwal, H.~L.{~}Bethlem, A.{~}Boeschoten, A.{~}Borschevsky, K.{~}Esajas, Y.{~}Hao \emph{et~al.},
\newblock Physical Review A {\bf 105},~052811 (2022).

\bibitem{oleynichenko2020towards}
A.~V.{~}Oleynichenko, A.{~}Zaitsevskii, and E.{~}Eliav,
\newblock Towards high performance relativistic electronic structure modelling: The exp-t program package,
\newblock in {\em Russian Supercomputing Days}, pp. 375--386, Springer, 2020.

\bibitem{Oleynichenko:website}
A.~V.{~}Oleynichenko, A.{~}Zaitsevskii, and E.{~}Eliav,
\newblock Exp-t, an extensible code for fock-space relativistic coupled cluster calculations,
\newblock \url{http://www.qchem.pnpi.spb.ru/expt}.

\bibitem{huber2013}
K.{~}Huber,
\newblock {\em Molecular spectra and molecular structure: IV. Constants of diatomic molecules},
\newblock Springer Science \& Business Media, 2013.

\bibitem{mosyagin2000comparison}
N.{~}Mosyagin, E.{~}Eliav, A.{~}Titov, and U.{~}Kaldor,
\newblock Journal of Physics B: Atomic, Molecular and Optical Physics {\bf 33},~667 (2000).

\bibitem{kaygorodov2021electron}
M.{~}Kaygorodov, L.{~}Skripnikov, I.{~}Tupitsyn, E.{~}Eliav, Y.{~}Kozhedub, A.{~}Malyshev, A.{~}Oleynichenko, V.{~}Shabaev, A.{~}Titov, and A.{~}Zaitsevskii,
\newblock Physical Review A {\bf 104},~012819 (2021).

\bibitem{kaygorodov2022ionization}
M.{~}Kaygorodov, D.{~}Usov, E.{~}Eliav, Y.{~}Kozhedub, A.{~}Malyshev, A.{~}Oleynichenko, V.{~}Shabaev, L.{~}Skripnikov, A.{~}Titov, I.{~}Tupitsyn \emph{et~al.},
\newblock Physical Review A {\bf 105},~062805 (2022).

\bibitem{kristiansen1986part1}
P.{~}Kristiansen and L.{~}Veseth,
\newblock The Journal of chemical physics {\bf 84},~2711 (1986).

\bibitem{chong1991theoretical}
D.~P.{~}Chong, S.~R.{~}Langhoff, and C.~W.{~}Bauschlicher~Jr,
\newblock The Journal of chemical physics {\bf 94},~3700 (1991).

\bibitem{leopold1987}
K.~R.{~}Leopold, K.~M.{~}Evenson, E.~R.{~}Comben, and J.~M.{~}Brown,
\newblock Journal of molecular spectroscopy {\bf 122},~440 (1987).

\bibitem{drouin2013isotopic}
B.~J.{~}Drouin,
\newblock The Journal of Physical Chemistry A {\bf 117},~10076 (2013).

\bibitem{kristiansen1986part2}
P.{~}Kristiansen and L.{~}Veseth,
\newblock The Journal of chemical physics {\bf 84},~6336 (1986).

\bibitem{raab1981precision}
F.{~}Raab, T.{~}Bergeman, D.{~}Lieberman, and H.{~}Metcalf,
\newblock Physical Review A {\bf 24},~3120 (1981).

\bibitem{ter1983determination}
J.{~}Ter~Meulen, W.{~}Majewski, W.~L.{~}Meerts, and A.{~}Dymanus,
\newblock Chemical Physics Letters {\bf 94},~25 (1983).

\bibitem{ter1986observation}
J.{~}Ter~Meulen, W.{~}Ubachs, and A.{~}Dumanus,
\newblock Chemical physics letters {\bf 129},~533 (1986).

\bibitem{fast2018precision}
A.{~}Fast, J.~E.{~}Furneaux, and S.~A.{~}Meek,
\newblock Physical Review A {\bf 98},~052511 (2018).

\bibitem{german1976high}
K.{~}German,
\newblock The Journal of Chemical Physics {\bf 64},~4192 (1976).

\bibitem{xin2003energy}
J.{~}Xin, I.{~}Ionescu, D.{~}Kuffel, and S.~A.{~}Reid,
\newblock Chemical physics {\bf 291},~61 (2003).

\bibitem{green1973dipole}
S.{~}Green,
\newblock The Journal of Chemical Physics {\bf 58},~4327 (1973).

\bibitem{coxon1980}
J.{~}Coxon,
\newblock Canadian Journal of Physics {\bf 58},~933 (1980).

\end{thebibliography}


\end{document}